# Towards Loosely Coupled Programming on Petascale Systems

Ioan Raicu[*], Zhao Zhang[+], Mike Wilde[#+], Ian Foster[#*+], Pete Beckman[#], Kamil Iskra[#], Ben Clifford[+]
[*]Department of Computer Science, University of Chicago, Chicago, IL, USA
[+]Computation Institute, University of Chicago and Argonne National Laboratory, Chicago, IL, USA
[#]Mathematics and Computer Science Division, Argonne National Laboratory, Argonne IL, USA
iraicu@cs.uchicago.edu, zhaozhang@uchicago.edu, {wilde,foster,beckman,iskra}@mcs.anl.gov, benc@ci.uchicago.edu

***Abstract*— We have extended the Falkon lightweight task execution framework to make loosely coupled programming on petascale systems a practical and useful programming model. This work studies and measures the performance factors involved in applying this approach to enable the use of petascale systems by a broader user community, and with greater ease. Our work enables the execution of highly parallel computations composed of loosely coupled serial jobs with no modifications to the respective applications. This approach allows a new—and potentially far larger—class of applications to leverage petascale systems, such as the IBM Blue Gene/P supercomputer. We present the challenges of I/O performance encountered in making this model practical, and show results using both microbenchmarks and real applications from two domains: economic energy modeling and molecular dynamics. Our benchmarks show that we can scale up to 160K processor-cores with high efficiency, and can achieve sustained execution rates of thousands of tasks per second.**

## I. Introduction

Emerging petascale computing systems, such as IBM's Blue Gene/P [1], incorporate high-speed, low-latency interconnects and other features designed to support tightly coupled parallel computations. Most of the applications run on these computers have a single program multiple data (SMPD) structure, and are commonly implemented by using the Message Passing Interface (MPI) to achieve the needed inter-process communication.

We want to enable the use of these systems for task-parallel applications, which are linked into useful workflows through the looser task-coupling model of passing data via files between dependent tasks. This potentially larger class of task-parallel applications is precluded from leveraging the increasing power of modern parallel systems because the lack of efficient support in those systems for the "scripting" programming model [2]. With advances in e-Science and the growing complexity of scientific analyses, more scientists and researchers rely on various forms of scripting to automate end-to-end application processes involving task coordination, provenance tracking, and bookkeeping. Their approaches are typically based on a model of loosely coupled computation, in which data is exchanged among tasks via files, databases or XML documents, or a combination of these. Vast increases in data volume combined with the growing complexity of data analysis procedures and algorithms have rendered traditional manual processing and exploration unfavorable as compared with modern high performance computing processes automated by scientific workflow systems. [3]

The problem space can be partitioned into four main categories (Figure 1). 1) At the low end of the spectrum (low number of tasks and small input size), we have tightly coupled MPI applications (white). 2) As the data size increases, we move into the analytics category, such as data mining and analysis (blue); MapReduce [4] is an example for this category. 3) *Keeping data size modest, but increasing the number of tasks moves us into the loosely coupled applications involving many tasks (yellow)*; Swift/Falkon [5, 6] and Pegasus/DAGMan [7] are examples of this category. 4) Finally, the combination of both many tasks and large datasets moves us into the data-intensive many task computing category (green); examples of this category are Swift/Falkon and data diffusion [8], Dryad [9], and Sawzall [10]. This paper focuses on the third category, at the largest scales of today's supercomputers on hundreds of thousands of processors.

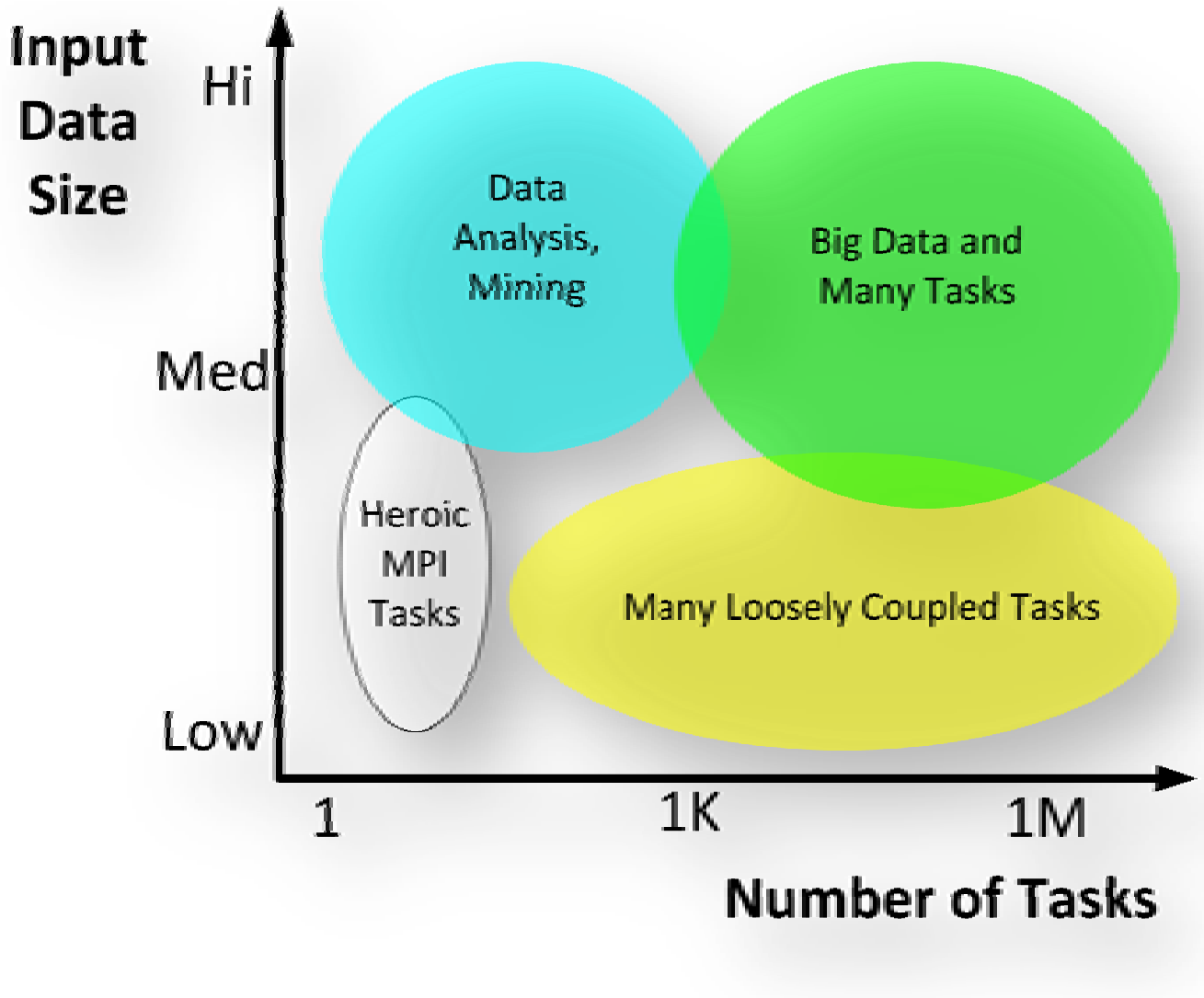

Figure 1: Problem types with respect to data size and number of tasks

This work was supported in part by the NASA Ames Research Center GSRP grant number NNA06CB89H, the Mathematical, Information, and Computational Sciences Division subprogram of the Office of Advanced Scientific Computing Research, Office of Science, U.S. Dept. of Energy, under Contract DE-AC02-06CH11357, and the National Science Foundation under grant OCI-0721939.

### A. *Many-Task Computing (MTC)*

Grids have been the preferred platform for loosely coupled applications that tend to be managed and executed through workflow systems or parallel programming systems. These loosely coupled applications make up a new class of applications called Many-Task Computing (MTC), which are composed of many tasks (both independent and dependent tasks) that can be individually scheduled on many different computing resources across multiple administrative boundaries to achieve some larger application goal. MTC is reminiscent of high throughput computing (HTC); MTC differs from HTC, however, in the emphasis on using much large numbers of computing resources over short periods of time to accomplish many computational tasks, where the primary metrics are in seconds (e.g., FLOPS, tasks/sec, MB/sec I/O rates). HTC, on the other hand, requires large amounts of computing for longer times (months and years, rather than hours and days), where the primary metrics are generally in operations per month) [11].

MTC denotes high-performance computations comprising multiple distinct activities, coupled via file system operations or message passing. Tasks may be small or large, uniprocessor or multiprocessor, compute-intensive or data-intensive. The set of tasks may be static or dynamic, homogeneous or heterogeneous, loosely coupled or tightly coupled. The aggregate number of tasks, quantity of computing, and volumes of data may be extremely large. Is MTC really different enough to justify coining a new term? There are certainly other choices we could have used instead, such as multiple program multiple data (MPMD), high throughput computing, workflows, capacity computing, or embarrassingly parallel.

MPMD is a variant of Flynn's original taxonomy [12], used to denote computations in which several programs each operate on different data at the same time. MPMD can be contrasted with SPMD, in which multiple instances of the same program each execute on different processors, operating on different data. MPMD lacks the emphasis that a set of tasks can vary dynamically. High throughput computing [11], a term coined by Miron Livny within the Condor project [13], to contrast workloads for which the key metric is not floating-point operations per second (as in high performance computing) but "per month or year." MTC applications are often just as concerned with performance as is the most demanding HPC application; they just don't happen to be SPMD programs. The term "workflow" was first used to denote sequences of tasks in business processes, but the term is sometimes used to denote any computation in which control and data passes from one "task" to another. We find it often used to describe many-task computations (or MPMD, HTC, MTC, etc.), making its use too general. "Embarrassingly parallel computing" is used to denote parallel computations in which each individual (often identical) task can execute without any significant communication with other tasks or with a file system. Some MTC applications will be simple and embarrassingly parallel, but others will be extremely complex and communication-intensive, interacting with other tasks and shared file-systems.

Is "many task computing" a useful distinction? Perhaps we could simply have said "applications that are communication-intensive but are not naturally expressed in MPI". Through the new term MTC, we drawing attention to the many computations that are heterogeneous but not "happily" parallel.

### B. *Hypothesis*

We claim that MTC applications can be executed efficiently on today's supercomputers; this paper provides empirical evidence to prove our hypothesis. The paper also describes the set of problems that must be overcome to make loosely coupled programming practical on emerging petascale architectures: local resource manager scalability and granularity, efficient utilization of the raw hardware, shared file system contention, and application scalability. We address these problems, and identify the remaining challenges that need to be overcome to make loosely coupled supercomputing a practical reality. Through our work, we have enabled a Blue Gene/P to efficiently support loosely coupled parallel programming without any modifications to the respective applications (except for recompilation), enabling the same applications that execute in a distributed Grid environment to be run efficiently on a supercomputer. The Blue Gene/P that we refer to in this paper is the new IBM Blue Gene/P supercomputer (also known as Intrepid) at the U.S. Department of Energy's Argonne National Laboratory, which is ranked number 3 in the Top500 rankings [15] with 160K processor-cores with a Rpeak of 557 TF and Rmax of 450 TF.

We validate our hypothesis by testing and measuring two systems, Swift [5, 14] and Falkon [6], which have been used to execute large-scale loosely coupled applications on clusters and Grids. We present results for both microbenchmarks and real applications executed on the Blue Gene/P. Microbenchmarks show that we can scale to 160K processor-cores with high efficiency, and can achieve sustained execution rates of thousands of tasks per second. We also investigated two applications from different domains, economic energy modeling and molecular dynamics, and show excellent application scalability, speedup and efficiency as they scale to 128K cores. Note that for the remainder of this paper, we will use the terms processors, CPUs, and cores interchangeably.

### C. *Why Petascale Systems for MTC Applications?*

One could ask, why use petascale systems for problems that might work well on terascale systems? We point out that petascale scale systems are more than just many processors with large peak petaflop ratings. They normally come well balanced, with proprietary, high-speed, and low-latency network interconnects to give tightly-coupled applications that use MPI good opportunities to scale well at full system scales. Even IBM has proposed in their internal project Kittyhawk [16] that the Blue Gene/P can be used to run non-traditional workloads, such as those found in the general Internet, which are by definition part of a loosely coupled system.

Four factors motivate the support of MTC applications on petascale HPC systems.

*1) The I/O subsystem of petascale systems offers unique capabilities needed by MTC applications.* For example, collective I/O operations could be implemented to use the specialized high-bandwidth and low-latency interconnects. We have not explored collective I/O operations in this work, but will do so in future work. MTC applications could be composed of individual tasks that are themselves parallel programs, many tasks operating on the same input data, and tasks that need considerable communication among themselves. Furthermore, the aggregate shared file system performance of a supercomputer can be potentially larger than that found in a distributed infrastructure (i.e., Grid), with data rates in the 8GB/s range, rather than the more typical 0.1GB/s to 1GB/s range of most Grid sites.

*2) The cost to manage and run on petascale systems like the Blue Gene/P is less than that of conventional clusters or Grids.* [16] For example, a single 13.9 TF Blue Gene/P rack draws 40 kilowatts, for 0.35 GF/watt. Two other systems that get good compute power per watt consumed are the SiCortex with 0.32 GF/watt and the Blue Gene/L with 0.23 GF/watt. In contrast, the average power consumption of the Top500 systems is 0.12 GF/watt [15]. Furthermore, we also argue that it is more cost effective to manage one large system in one physical location, rather than many smaller systems in geographically distributed locations.

*3)Large-scale systems inevitably have utilization issues.* Hence it is desirable to have a community of users who can leverage traditional back-filling strategies to run loosely coupled applications on idle portions of petascale systems.

*4) Perhaps most important, some applications are so demanding that only petascale systems have enough compute power to get results in a reasonable timeframe, or to leverage new opportunities. With petascale processing of ordinary applications, it becomes possible to perform vast computations quickly, thus answering questions in a timeframe that can make a quantitative difference in addressing significant scientific challenges or responding to emergencies.* This work has opened up the door for many important serial applications to use emerging petascale systems.

## II. Related Work

Only recently have parallel systems with 100K cores or more become available for open science research. Even scarcer is experience or success with loosely coupled programming at this scale. We found two papers [22, 23] that explored a similar space, focusing on HTC on the IBM Blue Gene/L [18].

Cope et al. [22] aimed at integrating their solution as much as possible in the Cobalt scheduling system (as opposed to bringing in another system such as Falkon); their implementation was on the Blue Gene/L using the HTC-mode [21] support in Cobalt, and the majority of the performance study was done at a small scale (64 nodes, 128 processors). The results of Cope et al. were at least one order of magnitude worse at small scales than the results we obtained in this paper, and the performance gap would only increase with larger-scale tests as their approach has higher overheads (i.e., nodes reboot after each task, in contrast with simply forking another process). Peter's et al. from IBM also recently published some performance numbers on the HTC-mode native support in Cobalt [23], which show a similar one order of magnitude difference between HTC-mode on Blue Gene/L and our Falkon support for MTC workloads on the Blue Gene/P. Subsection IV.C.1 compares and contrasts the performance between our proposed system on the Blue Gene/P and the results presented by Cope at al. [22] and Peters et al. [23].

In the world of high throughput computing, systems such as Condor [13], MapReduce [4], Hadoop [24], and BOINC [25] have used highly distributed pools of processors, but the focus of these systems has not been on single highly parallel machines such as those we focus on here. MapReduce is typically applied to a data model consisting of name/value pairs, processed at the programming language level. It has several similarities to the approach we apply here, in particular its ability to spread the processing of a large dataset to thousands of processors. However, it is far less amenable to the utilization and chaining of exiting application programs, and it often involves the development of custom filtering scripts. We have compared our work with Condor glide-ins [16] in the past [6], but our work focuses on performance and efficiency, while Condor emphasizes more on robustness and recoverability, which limits its efficiency for MTC applications in large-scale systems. An approach by Reid called "task farming" [26], also at the programming language level, has been evaluated on the Blue Gene/L as a proof of concept, but offered no performance evaluation for comparison, and required that applications be modified to run over the proposed middleware.

## III. Requirements and Implementation

The contribution of this work is the ability to enable a new class of applications—large-scale and loosely coupled—to efficiently execute on petascale systems, which are traditionally HPC systems. This is accomplished primarily through three mechanisms: 1) *multi-level scheduling*, 2) *efficient task dispatch*, and 3) *extensive use of caching to minimize shared infrastructure* (e.g. file systems and interconnects).

*Multi-level scheduling* is essential on a system such as the Blue Gene/P because the local resource manager (LRM, Cobalt [27]) works at a granularity of psets [28], rather than individual computing nodes or processor cores. On the Blue Gene/P, a pset is a group of 64 quad-core compute nodes and one I/O node. Psets must be allocated in their entirety to user application jobs by the LRM, which imposes the constraint that the applications must make use of all 256 cores. Tightly coupled MPI applications are well suited for this constraint, but loosely coupled applications generally have many single processor jobs, each with possibly unique executables and parameters. Naively running such applications on the Blue Gene/P using the system's Cobalt LRM would yield a utilization of 1/256. We use multi-level scheduling to allocate compute resources from Cobalt at the pset granularity, and then make these resources available to applications at a single processor core granularity. Using this multi-level scheduling mechanism, we are able to launch a unique application, or the same application with unique arguments, on each core, and to launch such tasks repetitively throughout the allocation period. This capability is made possible through Falkon [6] and its resource provisioning mechanisms.

A related obstacle to loosely coupled programming when using the native Blue Gene/P LRM is the overhead of scheduling and starting resources. The Blue Gene/P compute nodes are powered off when not in use and must be booted when allocated to a job. As the compute nodes do not have local disks, the boot-up process involves reading the lightweight IBM compute node kernel (or Linux-based ZeptoOS [29] kernel image) from a shared file system, which can be expensive if compute nodes are allocated and de-allocated frequently. Using multi-level scheduling allows this high initial cost to be amortized over many jobs, reducing it to an insignificant overhead. With the use of multi-level scheduling, executing a job is reduced to its bare and lightweight essentials: loading the application into memory, executing it, and returning its exit code – a process that can occur in milliseconds. Contrast this with the cost of rebooting compute nodes, which is on the order of multiple seconds (for a single node) and can be as high as a thousand seconds in the case of concurrently booting 40K nodes (see Figure 3).

The second mechanism that enables loosely coupled applications to be executed on the Blue Gene/P is a *streamlined task submission* framework (Falkon [6]). Falkon relies on LRMs for many functions (e.g., reservation, policy-based scheduling, accounting) and client frameworks such as workflow systems or distributed scripting systems for others (e.g., recovery, data staging, job dependency management). This specialization allows it to achieve several orders of magnitude higher performance (2534 tasks/sec in a Linux cluster environment, 3186 tasks/sec on the SiCortex, and 3071 tasks/sec on the Blue Gene/P, compared to 0.5 to 22 jobs per second for traditional LRMs such as Condor [13] and PBS [30] – see section IV.C). These high throughputs are critical in running large number of tasks on many processors as efficiently as possible. For example, running many 60-second tasks on 160K processors on the Blue Gene/P requires that we sustain an average throughput of 2730 tasks/sec; considering the best LRM performance of 22 tasks/sec [31], we would need two hour long tasks to get reasonable efficiency.

The third mechanism we employ for enabling loosely coupled applications to execute efficiently on the Blue Gene/P is *extensive caching* of application data to allow better application scalability by avoiding shared file systems. Since workflow systems frequently employ files as the primary communication medium between data-dependent jobs, having efficient mechanisms to read and write files is critical. The compute nodes on the Blue Gene/P do not have local disks, but they have both a shared file system (GPFS [32]) and local file system implemented in RAM ("ramdisk"). We make extensive use of the ramdisk local file system, to cache files such as application scripts and binary executables, static input data that is constant across many jobs running an application, and in some cases output data from the application until enough data is collected to allow efficient writes to the shared file system. We found that naively executing applications directly against GPFS yielded unacceptably poor performance, but with successive levels of caching we were able to increase the execution efficiency to within a few percent of ideal.

The caching we refer to in this work is a different mechanism from the data diffusion described in previous work [34, 35, 8]. Data diffusion deals with dynamic data caching and replication, as well as with data-aware scheduling. Because of the network topology of the Blue Gene/P, and the architecture changes in which we distributed the Falkon dispatcher (see Section III.B), where compute nodes are grouped into private networks per pset (in groups of 256 CPUs) using the Tree network, we have not been able to use data diffusion in its current form on the Blue Gene/P. We have made good progress in implementing TCP/IP over MPI to enable the use of the Torus network for node-to-node communication, which should allow us to test data diffusion on the BG/P; we will discuss this more in the future work section. On the other hand, the simple caching scheme we have employed on the Blue Gene/P deals with two kinds of data: 1) static data (application binaries, libraries, and common input data) that is cached at all compute nodes, and the caches are reused for each task; and 2) dynamic data (input data specific for a single task) that is cached on one compute node, and tasks can run completely local in both reading and writing data, and finally persisting the cache contents to a shared file system. Note that dynamic data is only used once by one task, and needs to be transferred from the persistent storage location again if another task needs the same input data. This simple caching scheme has proved to be quite effective in scaling applications up to 128K processors, while the same applications and workloads didn't scale well beyond 8K processors. Our caching strategy is completely automated, via a wrapper script around the application.

### A. Swift and Falkon

To harness a wide array of loosely coupled applications that have already been implemented and executed in clusters and grids, we build on the Swift [5, 36] and Falkon [6] systems. Swift enables scientific workflows through a data-flow-based functional parallel programming model. It is a parallel scripting tool for rapid and reliable specification, execution, and management of large-scale science and engineering workflows. The runtime system in Swift relies on the CoG Karajan [33] workflow engine for efficient scheduling and load balancing, and it integrates with the Falkon light-weight task execution dispatcher for optimized task throughput and efficiency.

Swift and Falkon have been used in a variety of environments from clusters, to multi-site Grids (e.g., Open Science Grid [37], TeraGrid [38]), to specialized large machines (SiCortex [39]), to supercomputers (e.g., Blue Gene/P [1]). Large-scale applications from many domains (e.g., astronomy [40, 6], medicine [41, 6, 42], chemistry [36], molecular dynamics [43], and economics [44, 45, 45]) have been run at scales of up to millions of tasks on up to hundreds of thousands of processors.

### B. Implementation Details

Significant engineering efforts were needed to get Falkon to work on systems such as the Blue Gene/P; this subsection discusses these extensions.

***Static Resource Provisioning:*** When using static resource provisioning, an application requests a number of processors for a fixed duration directly from the Cobalt LRM. For example, the command "*falkon-start-bgp-ram.sh prod 1024 60*" submits a single job to Cobalt to the "prod" queue and asks for 1024 nodes (4096 processors) for 60 minutes; once the job goes into a running state and the Falkon framework is bootstrapped, the application interacts directly with Falkon to submit single processor tasks for the duration of the allocation.

***Alternative Implementations:*** Performance depends critically on the behavior of our task dispatch mechanisms. The initial Falkon implementation was 100% Java, and made use of GT4 Java WS-Core to handle Web Services communications. [46] The Java-only implementation works well in typical Linux clusters and Grids; but the lack of Java on the Blue Gene/L, Blue Gene/P compute nodes, and SiCortex prompted us to re-implement some functionality in C. To keep the implementation simple yet able to support these specialized systems, we used a simple TCP-based protocol (to replace the prior WS-based protocol), internally between the dispatcher and the executor. We implemented a new component called TCPCore to handle the TCP-based communication protocol. TCPCore is a component to manage a pool of threads that lives in the same JVM as the Falkon dispatcher, and uses in-memory notifications and shared objects for communication. For performance reasons, we implemented persistent TCP sockets so connections can be reused across tasks..

***Distributed Falkon Architecture:*** The original Falkon architecture [6] used a single dispatcher (running on one login node) to manage many executors (running on compute nodes). The architecture of the Blue Gene/P is hierarchical, in which there are 10 login nodes, 640 I/O nodes, and 40K compute nodes. This led us to the offloading of the dispatcher from one login node (quad-core 2.5GHz PPC) to the many I/O nodes (quad-core 0.85GHz PPC); Figure 2 shows the distribution of components on different parts of the Blue Gene/P.

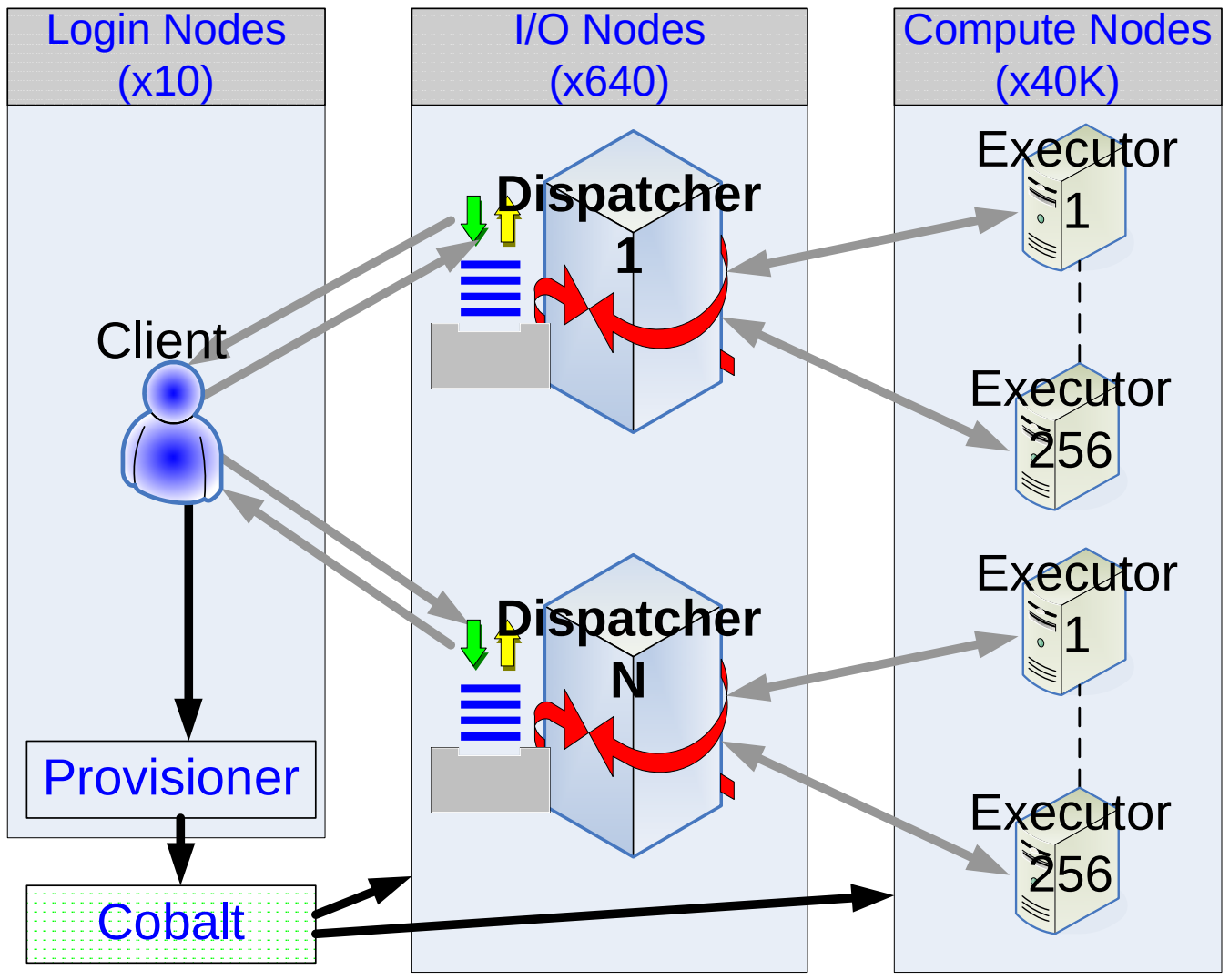

Figure 2: 3-Tier Falkon Architecture for BG/P.

Experiments show that a single dispatcher, when running on modern node with 4 to 8 cores at 2GHz+ and 2GB+ of memory, can handle thousands of tasks per second and tens of thousands of executors. As we ramped up our experiments to 160K processors (each executor running on one processor), however the centralized design began to show its limitations. One limitation (for scalability) was the fact that our implementation maintained persistent sockets to all executors (two sockets per executor). With the current implementation, we had trouble scaling a single dispatcher to 160K executors (320K sockets). Another motivation for distributing the dispatcher was to reduce the load on login nodes. The system administrators of the Blue Gene/P did not approve of the high system utilization (both memory and processors) of a login node for extended periods of time when we were running intense MTC applications.

Our change in architecture from a centralized one to a distributed one allowed each dispatcher to manage a disjoint set of 256 executors, without requiring any inter-dispatcher communication. The most challenging architecture change was the additional client-side functionality to communicate and load balance task submission across many dispatchers, and to ensure that it did not overcommit tasks that could cause some dispatchers to be underutilized while others queued up tasks. Our new architecture solved both the scalability problem to 160K processors and the excess load on the login nodes.

***Reliability Issues at Large Scale:*** We discuss reliability only briefly, to explain how our approach addresses this critical requirement. The Blue Gene/L has a mean-time-to-failure (MTBF) of 10 days [18], which can pose challenges for long-running applications. When running loosely coupled applications via Swift and Falkon, the failure of a single node affects only the task(s) being executed by the failed node at the time of the failure. I/O node failures affect only their respective psets (256 processors); these failures are identified by heartbeat messages or communication failures. Falkon has mechanisms to identify specific errors and act on them with specific actions. Most errors are generally passed back up to the client (in this case, Swift) to deal with them, but other (known) errors can be handled by Falkon directly by rescheduling the tasks. Falkon can suspend offending nodes if too many tasks fail in a short period of time. Swift maintains persistent state that allows it to restart a parallel application script from the point of failure, re-executing only uncompleted tasks. There is no need for explicit check-pointing as is the case with MPI applications; check-pointing occurs inherently with every task that completes and is communicated back to Swift.

## IV. MICROBENCHMARKS PERFORMANCE

We use microbenchmarks to determine performance characteristics and potential bottlenecks on systems with many cores. We measure startup costs, task dispatch rates, and costs for various file system operations (read, read+write, invoking scripts, mkdir, etc.) on the shared file systems (GPFS) that we use when running large-scale applications.

### *A. System Descriptions*

The IBM Blue Gene/P supercomputer Intrepid [1, 47] (hosted at Argonne National Laboratory) has quad-core processors with a total of 160K cores. The Blue Gene/P is rated at 557TF Rmax (450TF Rpeak) with 160K PPC450 processors running at 850MHz, with a total of 80 TB of main memory. The Blue Gene/P GPFS is rated at 8GB/s. In our experiments, we use an alpha version of Argonne's Linux-based ZeptoOS [29] compute node kernel.

### *B. Startup Costs*

Our first micro-benchmark captures the incremental costs involved (see Figure 3) in booting the Blue Gene/P at various scales (red), starting the Falkon framework (green), and initializing the Falkon framework (blue). On a single pset (256 processors), it takes 125 seconds to prepare Falkon to process the first task; on the full 160K processors, it takes 1326 seconds. At the smallest scale, starting and initializing the Falkon framework constitutes 31% of the total time; but at large scales, the boot time starts to dominate and on 160K nodes the Falkon framework takes only 17% of total time. We examine where the 1090 seconds is spent when booting ZeptOS on 160K nodes. The largest part of this time (708 seconds) is spent mounting GPFS. The next big block of time (213 seconds) is spent sending the kernels and ramdisks to the compute and I/O nodes. Mounting NFS (to access system software) takes 55 seconds. Starting various services from NFS, such as SSH, takes 85 seconds. These costs account for over 97% of the 1090 seconds required to boot the Blue Gene/P.

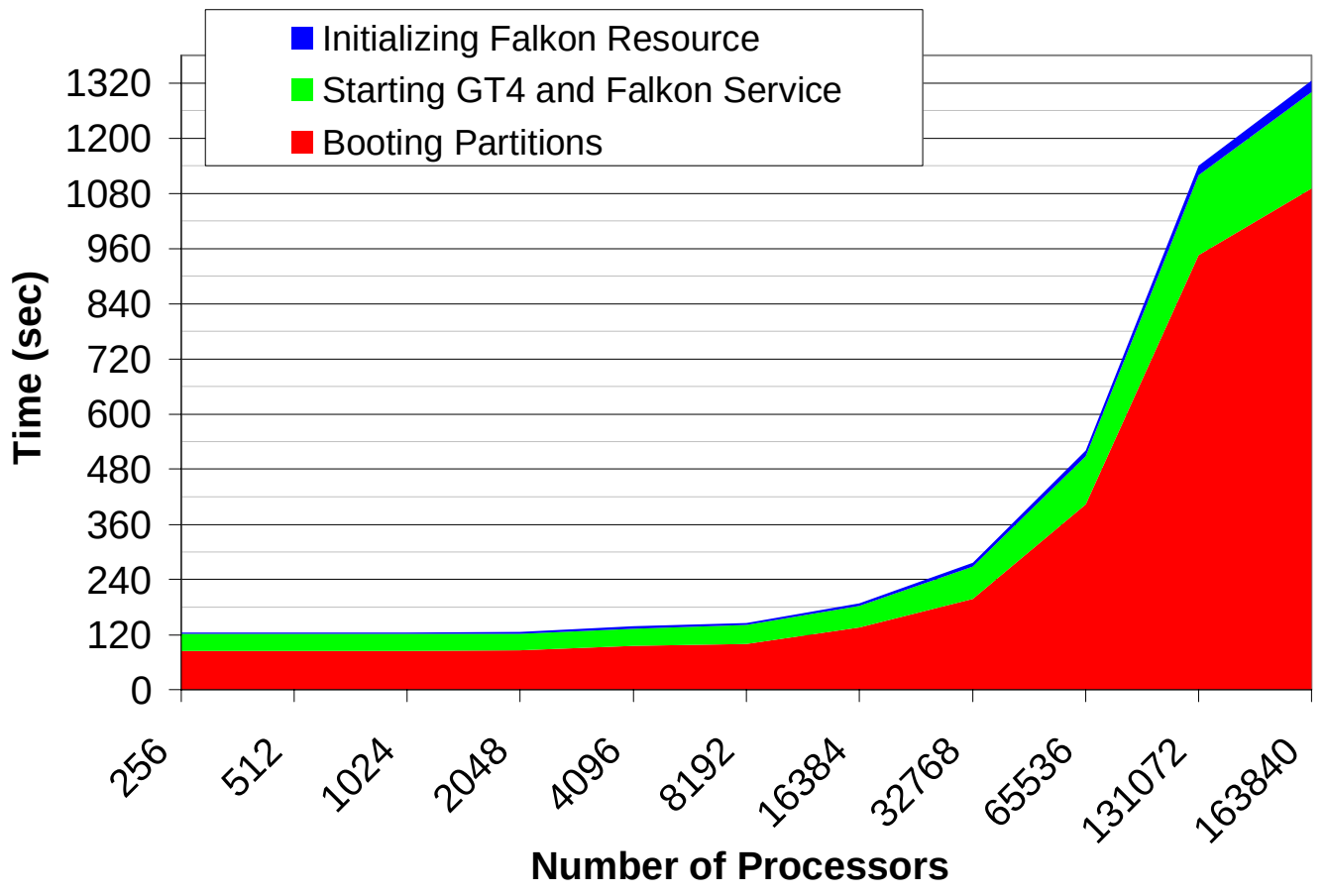

Figure 3: Startup costs in booting the Blue Gene/P, starting the Falkon framework, and initializing Falkon

## C. Falkon Task Dispatch Performance

One key component to achieving high utilization of large-scale systems is achieving high task dispatch and execute rates. Figure 4 shows the dispatch throughout of Falkon across various systems (Argonne/Univ. of Chicago Linux cluster, SiCortex, and Blue Gene/P) for both versions of the executor (Java and C, WS-based and TCP-based respectively) at significantly larger scales.

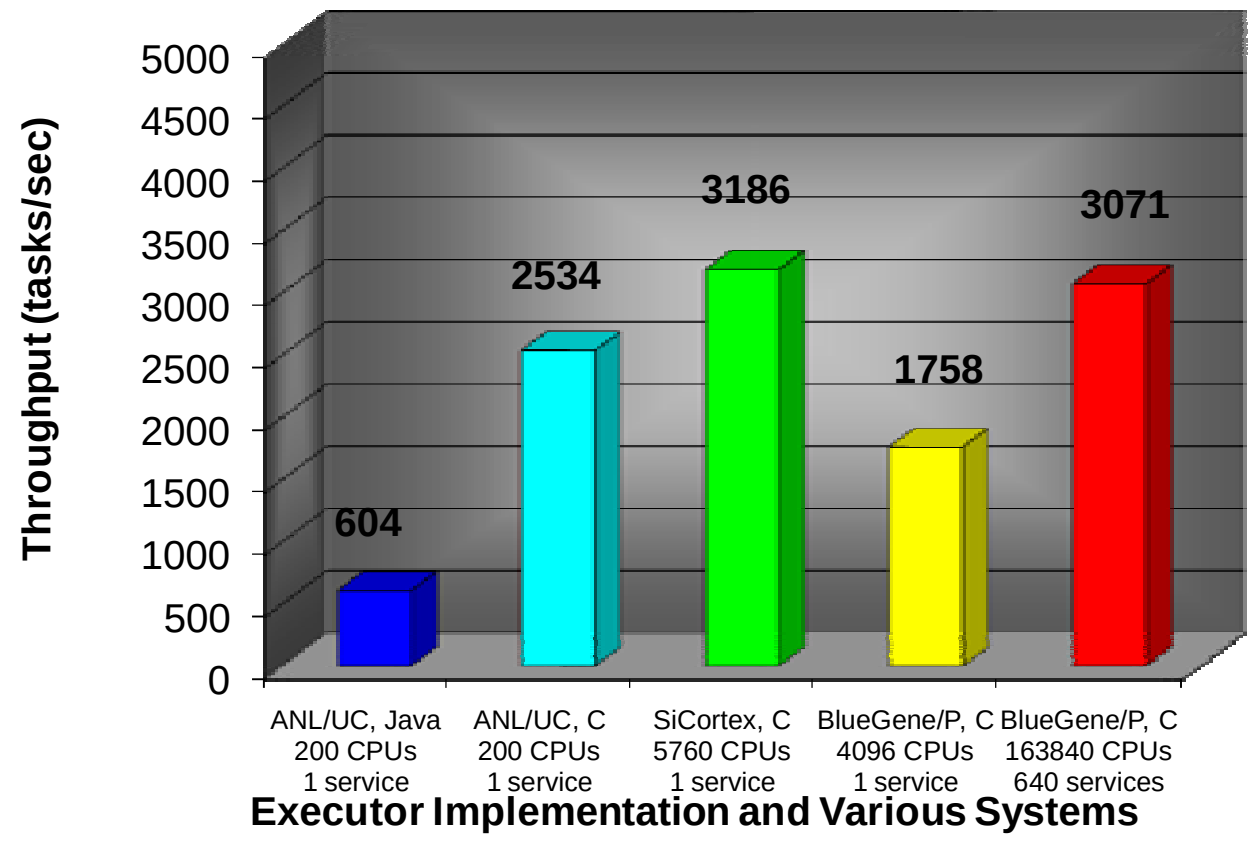

Figure 4: Falkon dispatch throughputs across various systems

In previous work [6] we reported that Falkon with a Java Executor and WS-based communication protocol achieves 487 tasks/sec in a Linux cluster (Argonne/Univ. of Chicago) with 256 CPUs, where each task was a “sleep 0” task with no I/O. Our latest benchmarks for the Java Executor on a faster machine achieved 604 tasks/sec and 2534 tasks/sec for the C Executor (Linux cluster, 1 dispatcher, 200 CPUs). The rest of the benchmarks only tested the C Executor as Java does not have good support on either the SiCortex or the Blue Gene/P; we achieved 3186 tasks/sec on the SiCortex (1 dispatcher, 5760 CPUs), 1758 tasks/sec on the Blue Gene/P with 1 dispatcher (4096 CPUs), and 3071 tasks/sec on the Blue Gene/P with 640 dispatchers (163840 CPUs). The throughput numbers that indicate “1 dispatcher” are tests done with the original centralized dispatcher running on a login node. The last throughput of 3071 tasks/sec was achieved with the dispatchers distributed over 640 I/O nodes, each managing 256 processors.

### 1) Comparing Falkon to Other LRMs and Solutions

It is instructive to compare task execution rates achieved by other local resource managers. In previous work [6], we measured Condor (v6.7.2, via MyCluster [11]) and PBS (v2.1.8) performance in a Linux environment (the same environment where we test Falkon and achieved 2534 tasks/sec throughputs). The throughput we measured for PBS was 0.45 tasks/sec and for Condor was 0.49 tasks/sec; other studies in the literature have measured Condor’s performance as high as 22 tasks/sec in a research prototype called Condor J2 [31].

We also tested the performance of Cobalt (the Blue Gene/P’s LRM), which yielded a throughput of 0.037 tasks/sec; recall that Cobalt also lacks the support for single processor tasks, unless HTC-mode [21] is used. HTC-mode means that the termination of a process does not release the allocated resource and initiates a node reboot, after which the launcher program is used to launch the next application. There is still some management required on the compute nodes, as exit codes from previous application invocations need to persist across reboots (e.g. to shared file system), be sent back to the client, and have the ability to launch an arbitrary application from the launcher program. Running Falkon in conjunction with Cobalt’s HTC-mode support yielded a 0.29 task/sec throughput. We investigated the performance of HTC-mode on the Blue Gene/L only at small scales, as we realized that it will not be sufficient for MTC applications because of the high overhead of node reboots across tasks; we did not pursue it at larger scales, or on the Blue Gene/P.

As we discussed in Section II, Cope et al. [22] also explored a similar space as we have, leveraging HTC-mode [21] support in Cobalt on the Blue Gene/L. The authors conducted various experiments, which we tried to replicate for comparison reasons. The authors measured an overhead of 46.4±21.2 seconds for running 60 second tasks on 1 pset of 64 processors on the Blue Gene/L. In a similar experiment running 64 second tasks on 1 pset of 256 processors on the Blue Gene/P, we achieve an overhead of 1.2±2.8 seconds, more than an order of magnitude better. Another comparison is the task startup time, which they measured to be on average about 25 seconds, but sometimes as high as 45 seconds; the startup times for tasks in our system are 0.8±2.7 seconds. Another comparison is average task load time by number of simultaneously submitted tasks on a single pset and executable image size of 8MB (tasks return immediately, so the reported run time shows overhead). The authors reported an average of 40~80 seconds for 32 simultaneous tasks on 32 compute nodes on the Blue Gene/L (1 pset, 64 CPUs). We measured our overheads of executing an 8MB binary to be 9.5±3.1 seconds on 64 compute nodes on the Blue Gene/P (1 pset, 256 CPUs). Since these times include the time it took to cache the binary in ramdisk, we believe these numbers will remain relatively stable as we scale up to full 160K processors. Note that the work by Cope et al. is based on Cobalt's HTC-mode [21], which implies that they perform a node reboot for every task, while we simply fork the application as a separate process for each task.

Peter's et al. also recently published some performance numbers on the HTC-mode native support in Cobalt [23]. Their results show a similar order of magnitude difference between the HTC-mode on Blue Gene/L and our Falkon support for MTC workloads on the Blue Gene/P. For example, the authors reported a workload of 32K tasks on 8K processors and noted that 32 dispatchers take 182.85 seconds to complete (an overhead of 5.58ms per task), but the same workload on the same number of processors using Falkon completed in 30.31 seconds with 32 dispatchers (an overhead of 0.92ms per task). Note that a similar workload of 1M tasks on 160K processors run by Falkon can be completed in 368 seconds, which translates to 0.35ms per task overhead.

### 2) *Efficiency and Speedup*

To better understand the performance achieved for different workloads, we measured performance as a function of task length. We made measurements in two different configurations: 1) 1 dispatcher and up to 2K processors, and 2) N/256 dispatchers on up to N=160K processors, with 1 dispatcher managing 256 processors. We varied the task lengths from 1 second to 256 seconds (using sleep tasks with no I/O), and ran workloads ranging from 1K tasks to 1M tasks (depending on the task lengths, to ensure that the experiments completed in a reasonable amount of time). Figure 5 shows the effects of efficiency of 1 dispatcher running on a faster login node (quad core 2.5GHz PPC) at relatively small scales. With 4 second tasks, we can get high efficiency (95%+) across the board (up to the measured 2K processors).

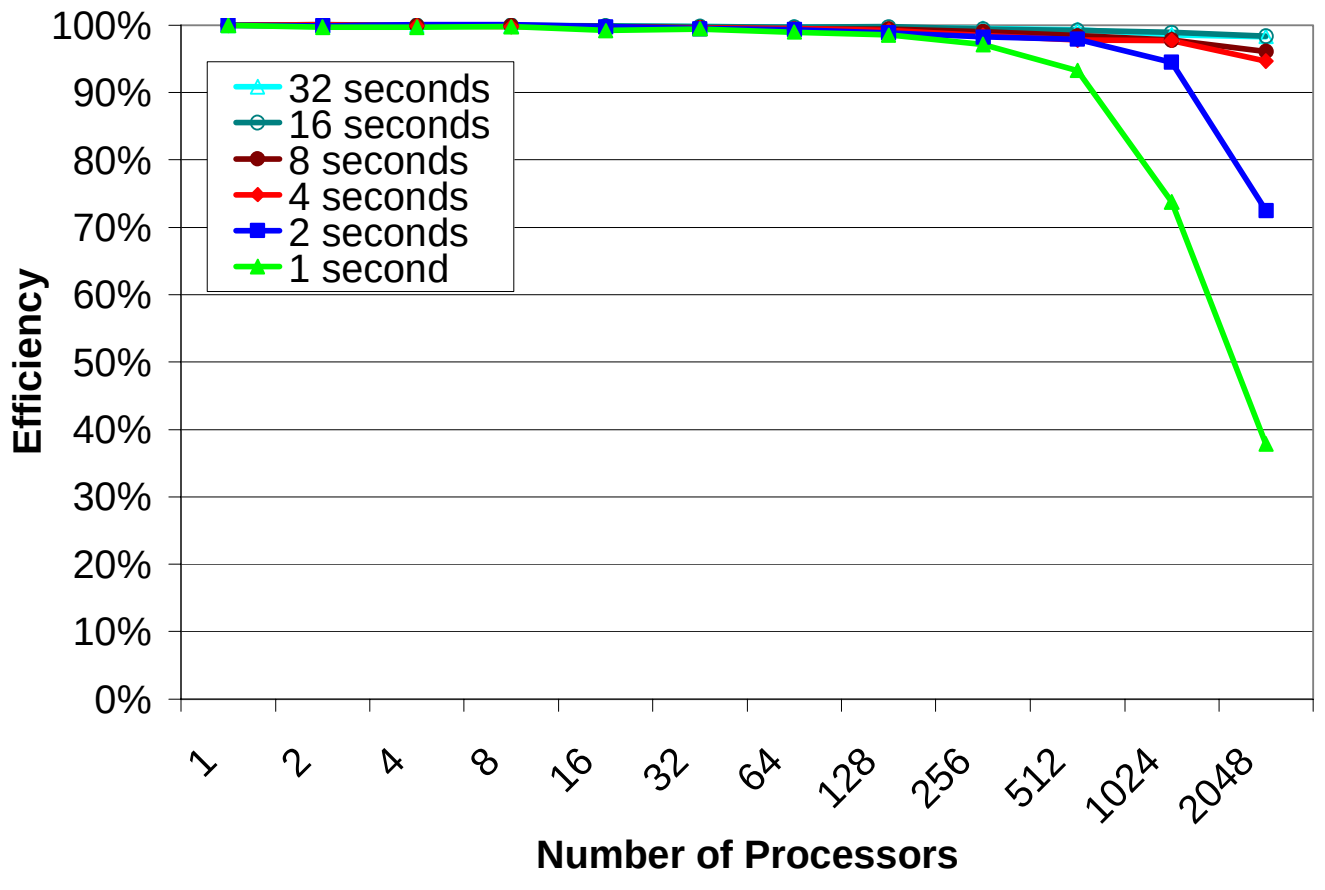

Figure 5: Efficiency graph for the Blue Gene/P for 1 to 2048 processors and task lengths from 1 to 32 seconds using a single dispatcher on a login node

Figure 6 shows the efficiency with the distributed dispatchers on the slower I/O nodes (quad core 850 MHz PPC) at larger scales. It is interesting to notice that the same 4 second tasks that offered high efficiency in the single dispatcher configuration now achieve relatively poor efficiency, starting at 65% and dropping to 7% at 160K processors. This is due to both the extra costs associated with running the dispatcher on slower hardware, and the increasing need for high throughputs at large scales. If we consider the 160K processor case, based on our experiments, we need tasks to be at least 64 seconds long to get 90%+ efficiency. Adding I/O to each task will further increase the minimum task length in order to achieve high efficiency.

To summarize: distributing the Falkon dispatcher from a single (fast) login node to many (slow) I/O nodes has both advantages and disadvantages. The advantage is that we achieve good scalability to 160K processors. The disadvantage is significantly worse efficiency at small scales (less than 4K processors) and short tasks (1 to 8 seconds). We believe both approaches are valid, depending on the application task execution distribution and scale of the application.

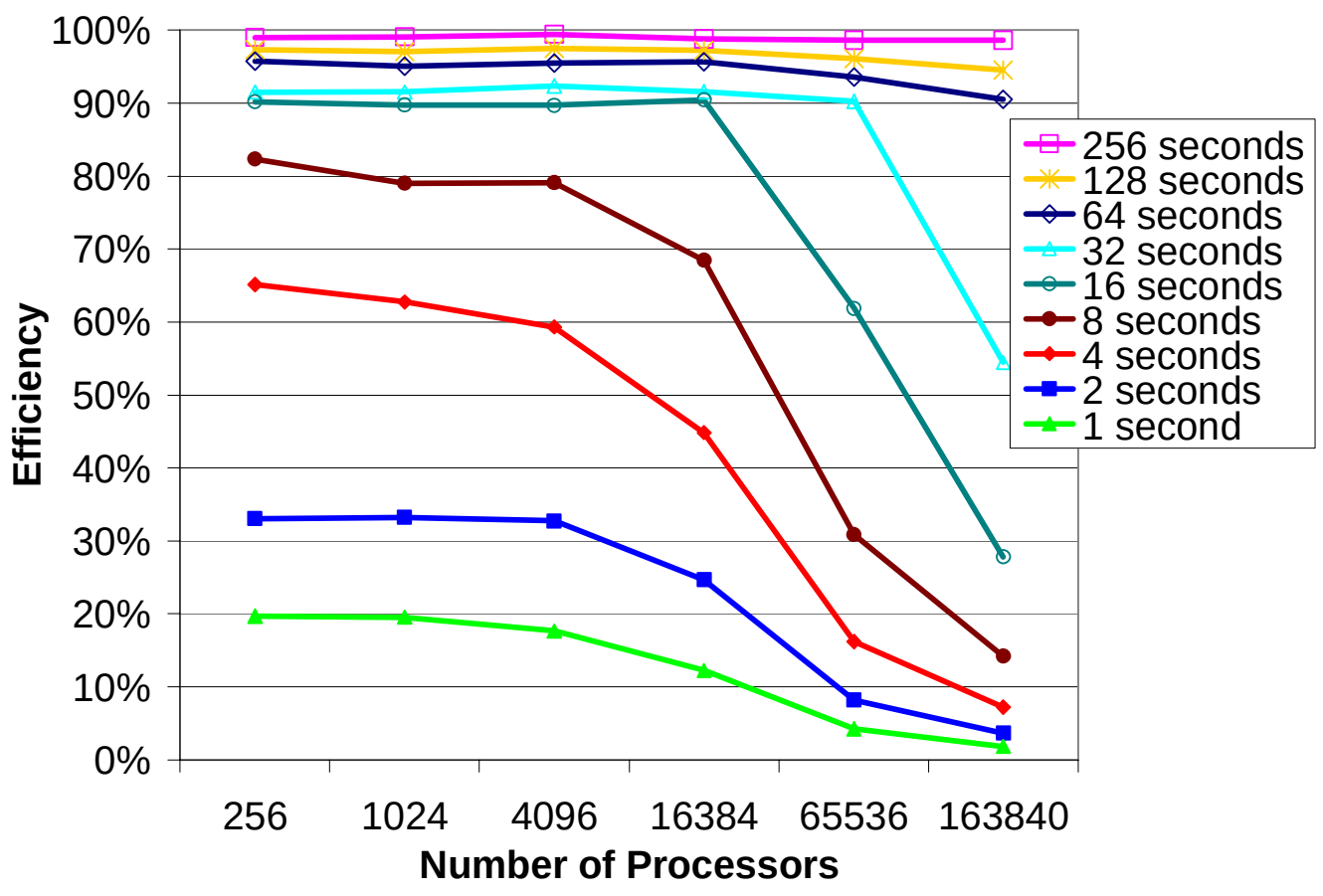

Figure 6: Efficiency graph for the Blue Gene/P for 256 to 160K processors and task lengths ranging from 1 to 256 seconds using N dispatchers with each dispatcher running on a separate I/O node

## D. *Shared File System Performance*

Another key component to getting high utilization and efficiency on large-scale systems is to understand the shared resources well. This sub-section discusses the shared file system performance of the Blue Gene/P. This performance is important because many MTC applications use files for inter-process communication, and these files are typically transferred from one node to another through the shared file system. Future work will remove this bottleneck, by using TCP pipes, MPI messages, or data diffusion [34, 8] to transfer files directly between compute nodes over the specialized networks of the Blue Gene/P.

We conducted several experiments (see Figure 7) with various data sizes (1KB to 10MB) on a varying number of processors (4 to 16K); we conducted both read-only tests (dotted lines) and read+write tests (solid lines). At 16K processors, we were not able to saturated GPFS – note the throughput lines never plateau. GPFS is configured with 16 I/O servers, each with 10Gb/s network connectivity, and can sustain 8GB/s aggregate I/O rates. We were able to achieve 4.4GB/s read rates, and 1.3GB/s read+write rates with 10MB files and 16K processors (we used the Linux "dd" utility to read or read+write data in 128KB blocks). We made our measurements in a production system, where the majority (90%+) of the system was in use by other applications, which might have been using the shared file system as well, influencing the results from this micro-benchmark.

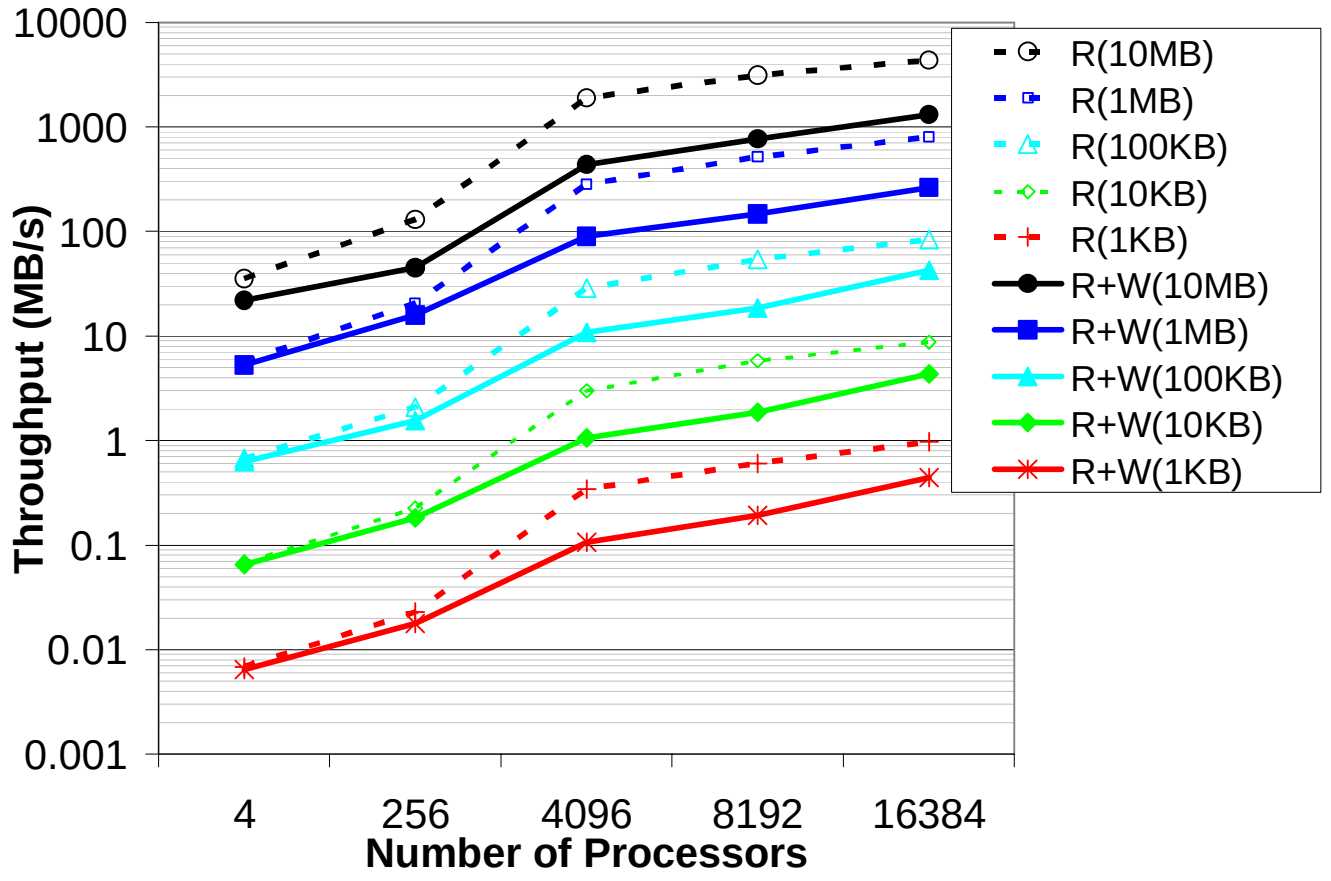

Figure 7: GPFS Throughput in MB/s measured through Falkon on various file sizes (1KB-10MB) and number of processors (4-16384)

It is important to understand how operation costs scale with increasing number of processors (see Figure 8). We tested file and directory creation in two scenarios: when all files or directories are created in the same directory (single dir), and when each file or directory is created in a unique pre-created directory (across many dirs). We investigated the costs to invoke a script from GPFS. We also measured the Falkon overhead of executing a trivial task with no I/O (sleep 0).

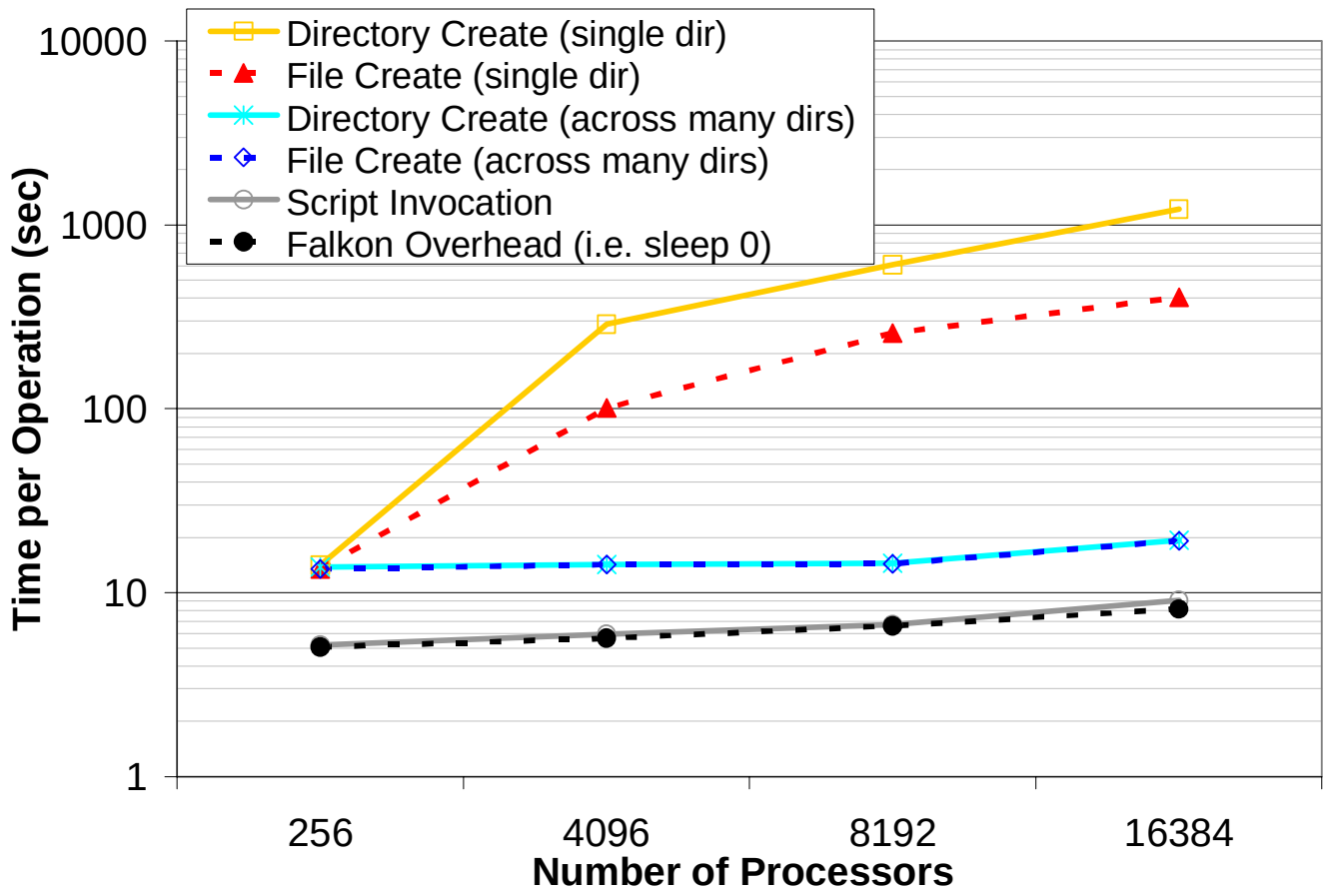

Figure 8: Time per operation (mkdir, touch, script execution) on GPFS on various number of processors (256-16384)

Both the file and directory create when performed in the same directory are expensive operations as we scale up the number of processors; for example, at 16K processors, it takes (on average) 404 seconds to create a file, and 1217 seconds to create a directory. These overheads translate to an aggregate throughput of 40 file creates per second and 13 directory creates per second. At these rates, 160K processors would require 68 and 210 minutes to create 160K files or directories. In contrast, when each file or directory create take place in a unique directory, performance is significantly improved; at small scales (256 processors), a file/directory create (in a unique directory) takes only 8 seconds longer than a basic task with no I/O; at large scales (16K processors), the overhead grows to 11 seconds. We conclude that I/O writes should be split over many directories, to avoid lock contention within GPFS from concurrent writers. These times reflect the costs of creating a file or directory when all processors perform the operation concurrently; many applications have a wide range of task lengths, and read/write operations occur only at the beginning and/or end of a task (as is the case with our caching mechanism), so the time per operation will be notably less because of the natural staggering of I/O calls.

## V. Loosely Coupled Applications

Synthetic tests and applications offer a great way to understand the performance characteristics of a particular system, but they do not always easily translate into predictions of how real applications with real I/O will behave. We have identified various loosely coupled applications as potential good candidates to run at large scales:

- *Ensemble runs to quantify climate model uncertainty*
- *Identify potential drug targets by screening a database of ligand structures against target proteins*
- *Study economic model sensitivity to parameters*
- *Analyze turbulence dataset from many perspectives*
- *Perform numerical optimization to determine optimal resource assignment in energy problems*
- *Mine collection of data from advanced light sources*
- *Construct databases of computed properties of chemical compounds*
- *Analyze data from the Large Hadron Collider*
- *Analyze log data from 100K-CPU parallel computations*

We use two applications (DOCK and MARS) to evaluate and demonstrate the utility of executing MTC applications on the Blue Gene/P.

### A. *Molecular Dynamics: DOCK*

This application, executed on the BG/P screens KEGG [48] compounds and drugs against important metabolic protein targets using the DOCK6 [43] application to simulate the "docking" of small molecules, or ligands, to the "active sites" of large macromolecules of known structure called "receptors" A compound that interacts strongly with a receptor (such as a protein molecule) associated with a disease may inhibit its function and thus act as a beneficial drug. The economic and health benefits of speeding drug development by rapidly screening for promising compounds and eliminating costly dead-ends is significant in terms of both resources and human life. In this application run, nine proteins that perform key enzymatic functions in the core metabolism of bacteria and humans were selected for screening against a database of 15,351 natural compounds and existing drugs in KEGG's ligand database.

*1) DOCK6 Performance Evaluation*

The binding affinity between each compound in the database and each protein was computed with 138,159 runs of DOCK6 on the Blue Gene/P. On 32 racks of the Blue Gene/P (128K cores at 0.85 GHz), these runs took 2807 seconds (see Figure 9), totaling 3.5 CPU years. The sustained utilization (while there were enough tasks to be done, roughly 600 seconds) was 95%, with the overall utilization being 30%. The large underutilization was caused by the heterogeneous task execution time (23/783/2802 +/- 300 seconds, for min/aver/max +/- stdev respectively). Expecting a significant underutilization, we had overlapped another application to start running as soon as the sustained period ended at around 600 seconds. The other application had enough work to be done that it actually used all of the idle CPUs from Figure 9 (the red area) with 97% utilization.

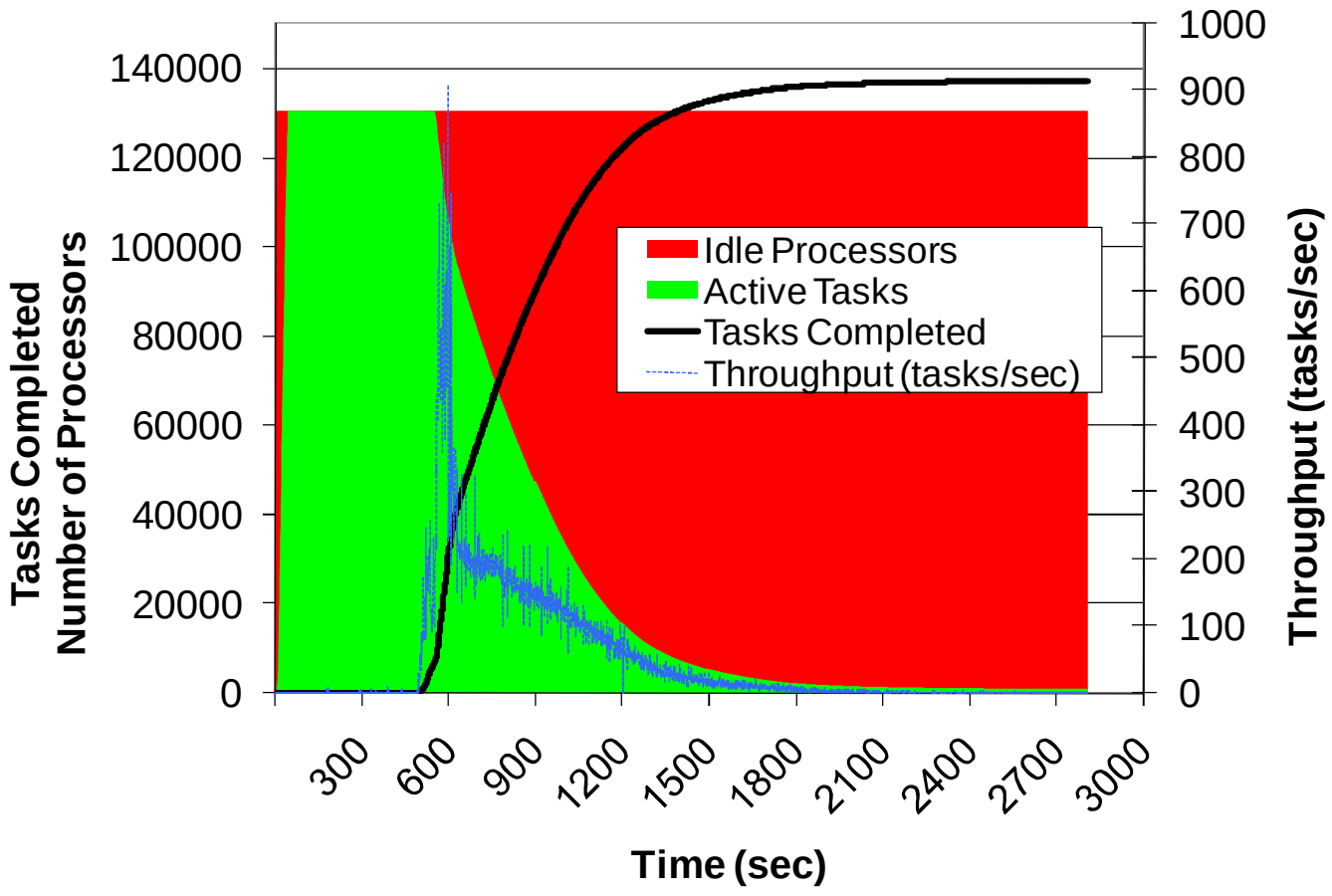

Figure 9: 138,159 DOCK6 runs on 131,072 CPU cores on Blue Gene/P

*2) DOCK5 Performance Evaluation*

We also worked with another group that had a larger set of runs using an older version of DOCK (version 5) [43]. This workload consisted of 934,803 molecules, which we ran on 116K CPU cores in 2.01 hours (see Figure 10). The per-task execution time was quite varied (even more so than the DOCK6 runs from Figure 9), with a minimum of 1 second, a maximum of 5030 seconds, and a mean of 713±560 seconds. The two-hour run has a sustained utilization of 99.6% (first 5700 seconds of experiment) and an overall utilization of 78% (due to the tail end of the experiment). Note that we had allocated 128K CPUs, but only 116K CPUs registered successfully and were available for the application run; the reason was the GPFS contention in bootstrapping Falkon on 32 racks, and was fixed in later large runs by moving the Falkon framework to RAM before starting, and by pre-creating log directories on GPFS to avoid lock contention. We have made dozens of runs at 32 and 40 rack scales, and we have not encountered this specific problem again.

Despite the loosely coupled nature of this application, our preliminary results show that the DOCK application performs and scales well to nearly full scale (116K of 160K CPUs). The excellent scalability (99.7% efficiency when compared to the same workload at half the scale of 64K CPUs) was achieved only after careful consideration was taken to avoid the shared file system, which included the caching of the multi-megabyte application binaries, and the caching of 35MB of static input data that would have otherwise been read from the shared file system for each job. Each job still had some minimal read and write operations to the shared file system, but they were on the order of tens of kilobytes (only at the beginning and end of computations), with the majority of the computations being in the hundreds of seconds, with an average of 713 seconds.

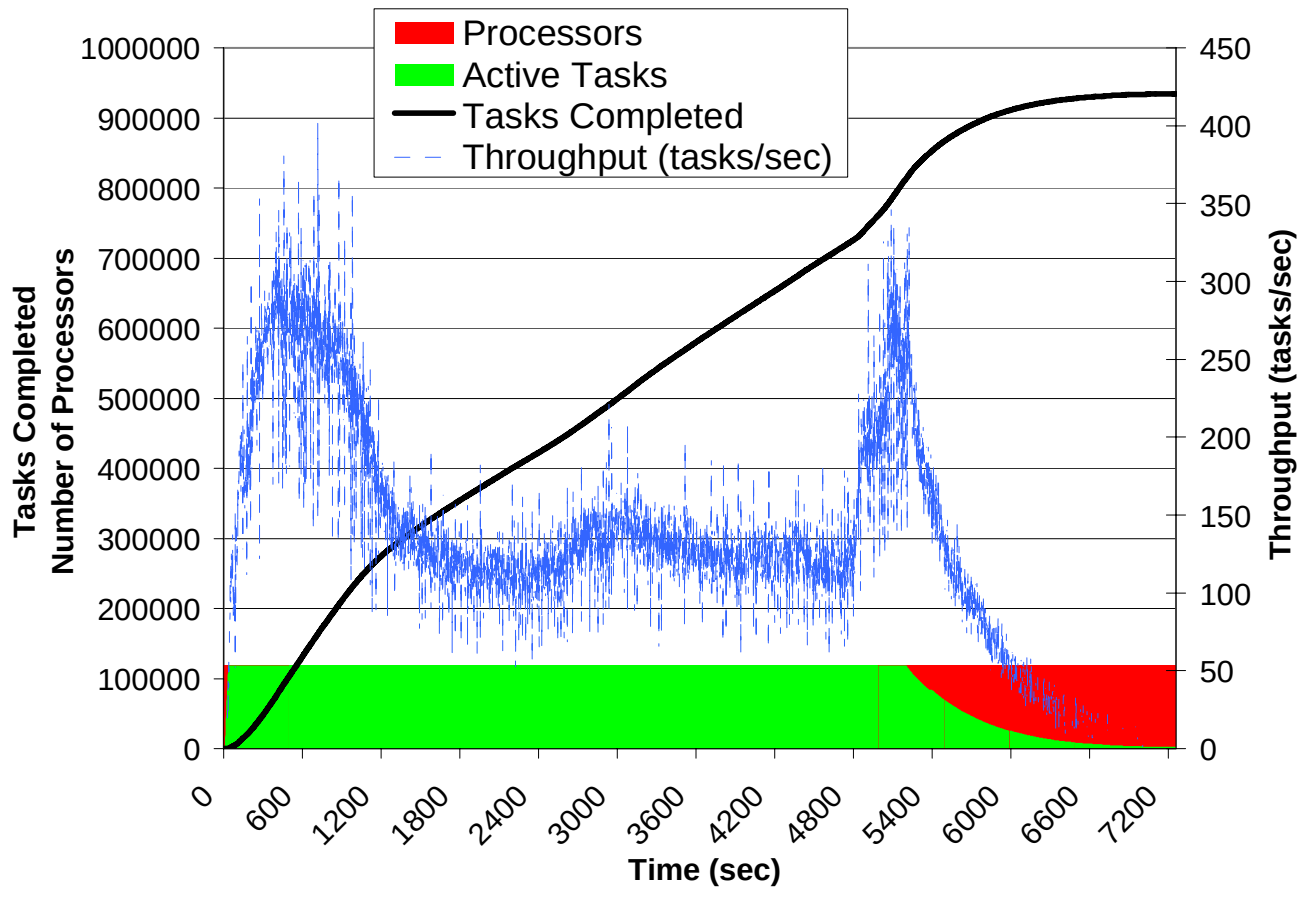

Figure 10: 934,803 DOCK5 runs on 118,784 CPU cores on Blue Gene/P

## B. *Economic Modeling: MARS*

The third application was MARS (Macro Analysis of Refinery Systems), an economic modeling application for petroleum refining developed by D. Hanson and J. Laitner at Argonne [44]. This modeling code performs a fast, broad-based simulation of the economic and environmental parameters of petroleum refining, covering over 20 primary and secondary refinery processes. MARS analyzes the processing stages for six grades of crude oil (from low-sulfur light to high-sulfur very-heavy and synthetic crude), as well as processes for upgrading heavy oils and oil sands. It analyses eight major refinery products including gasoline, diesel and jet fuel, and evaluates ranges of product shares. It models the economic and environmental impacts of the consumption of natural gas, the production and use of hydrogen, and coal-to-liquids co-production, and seeks to provide insights into how refineries can become more efficient through the capture of waste energy.

While MARS analyzes this large number of processes and variables, it does so at a coarse level. It consists of about 16K lines of C code, and can process many internal model execution iterations, with a range from 0.5 seconds (one iteration) to hours (many thousands of iterations) of Blue Gene/P CPU time. The goal of running MARS on the BG/P is to perform detailed multi-variable parameter studies of the behavior of all aspects of petroleum refining.

As a simple test of using the Blue Gene/P for refinery modeling, we performed a 2D parameter sweep to explore the sensitivity of the investment required to maintain petroleum production capacity, over a four-decade span, to variations in the diesel production yields from low sulfur light crude and medium sulfur heavy crude oils. This mimics one of the vast number of complex multivariate parameter studies that become possible with ample computing power. We then executed a far larger workload with 1M MARS tasks, combining both internal and external parameters sweeps, running eight MARS runs per processor core on 128K processors on the Blue Gene/P (see Figure 11).

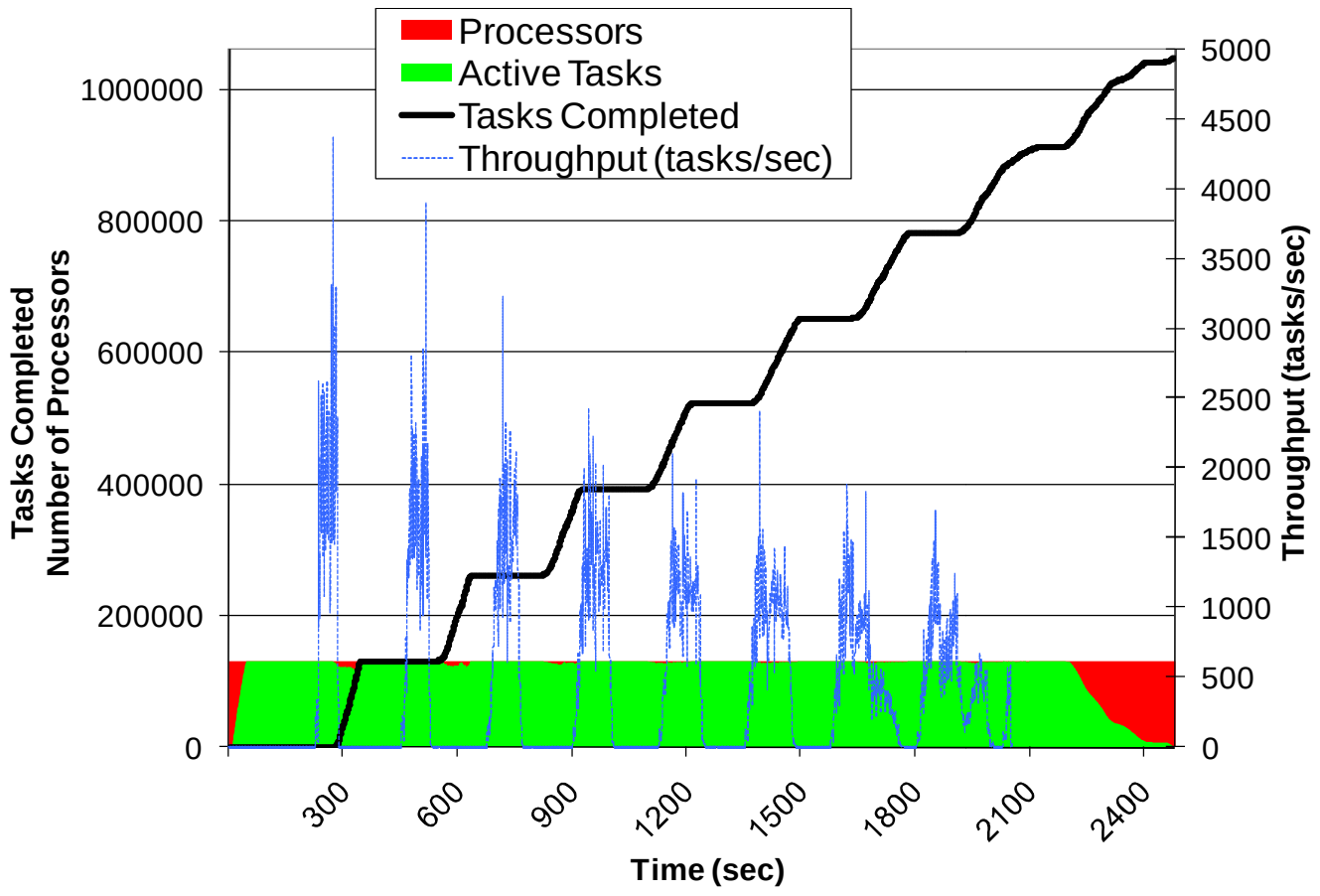

Figure 11: MARS application (summary view) on the Blue Gene/P; 1M tasks using 128K processor cores

The experiment consumed 9.3 CPU years and took 2483 seconds to complete, with an average of 280±10 seconds execution time per task. Our per task efficiency was 97%; but because of the time it took to dispatch the first wave of tasks, plus the time to ramp down the experiment, the overall efficiency of the experiment dropped to 88% with a speedup of 115,168X (ideal speedup being 130,816X).

### C. *Running applications through Swift*

The results presented in the preceding sections are from static workloads processed directly with Falkon. Swift, on the other hand, can be used to make workloads more dynamic, and reliable, and provide a natural flow from the results of an application to the input of the following stage in a more complex workflow. We ran a 16384 task workload for the MARS application on 8192 CPUs. Figure 12 shows the comparison of this workload side-by-side between the Falkon only run (green lines) and the Swift run (blue lines), which used Falkon as the execution engine; the dotted lines represent submitted tasks and the solid lines are the finished tasks.

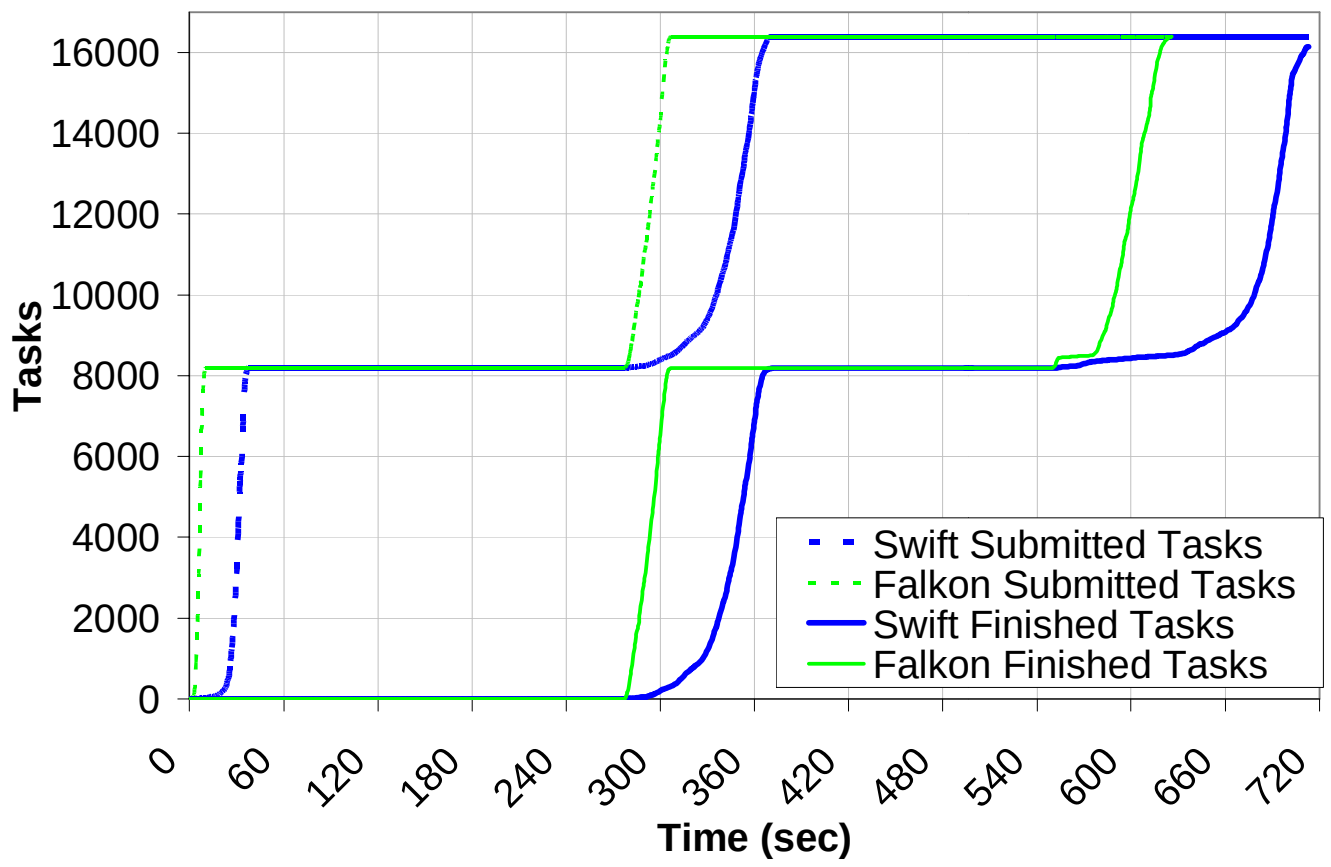

Figure 12: MARS application comparison between Swift and Falkon on the Blue Gene/P; 16K tasks using 8K processor cores

We see slower submission rates of Swift, as it takes 38 seconds to dispatch the first round of 8K tasks (as compared with 11 seconds of Falkon only). The transition between the first round of tasks and the second round also takes longer in the case of Swift with 94 seconds (as compared to 33 sec for Falkon only). Some of the additional time in this transition can be accounted to the larger standard deviation of task execution times (20 seconds for Swift as compared to 13 seconds for Falkon); other differences can be found in the additional overheads in Swift in processing each task. The ramp-down period was 161 seconds for Swift, but only took 77 seconds for Falkon. The end-to-end workload execution time was 713 seconds for Swift, and 626 seconds for Falkon, yielding an 88% efficiency. The per task efficiency was better with 91%, as it does not account for the ramp-up and ramp-down phase when not all processors are busy.

It was curious that we obtained higher average task execution times on 8K processors than on 128K processors (297±13 seconds on 8K processors and 280±10 seconds on 128K processors from Figure 11). The answer lies in the central role the shared file system has in loading the application in memory of the compute nodes, and reading and writing each task's input and output data. In our 128K processor experiment, we had dedicated access to the entire machine, so the shared file system load was all due to our own application data access patterns. In the 8K processor experiment, the rest of the 95% of the machine was busy running other applications, which in this case, slowed down our MARS application by 6% to 10% at the medium scale run of 8K processors.

Swift incurs its own overheads in addition to what Falkon experiences when running these applications. These overheads include 1) managing the data (staging data in and out, copying data from its original location to a workflow-specific location, and back from the workflow directory to the result archival location), 2) creating per-task working directories from the compute nodes (via mkdir on the shared file system), and 3) creating and tracking several status and log files for each task. Prior to achieving these good efficiency numbers (i.e. 88% efficiency on 8K processors), we were only able to achieve 20% efficiency using the default installation of Swift at the same scale. We investigated the main bottlenecks, which were determined to be from contention on the shared filesystem. We then applied three optimizations to Swift: 1) placing temporary directories in local ramdisk rather than the shared filesystem; 2) copying the input data to the local ramdisk of the compute node for each job execution; and 3) creating the per job logs on local ramdisk and copying them only to persistent shared storage at the completion of each job (rather than appending to a file on the shared file system at each job status change). These optimizations allowed us to increase the efficiency from 20% to 88% on 8192 processors for the MARS application. As future work, we will be working to scale up Swift to larger runs involving 100K to 1M tasks, and to maintain Swift's efficiency in the 90% range as we scale up to 160K processors.

## VI. CONCLUSIONS AND FUTURE WORK

Clusters with 50K+ processor cores (e.g., TACC Sun Constellation System, Ranger), Grids (e.g., TeraGrid) with over a dozen sites and 100K+ processors, and supercomputers with up to 256K processors (e.g., IBM Blue Gene/P) are now available to the scientific community. The effort described here has demonstrated the ability to manage and execute large-scale loosely coupled applications on petascale-class systems. These large HPC systems are considered efficient at executing tightly coupled parallel jobs within a particular machine using MPI to achieve inter-process communication. We proposed using HPC systems for loosely-coupled applications, which involve the execution of independent, sequential jobs that can be individually scheduled, and using files for inter-process communication. Our work shows that today's existing HPC systems are a viable platform to host MTC applications. We identified challenges in running these novel workloads on petascale systems, which can hamper the efficiency and utilization of these large-scale machines. These challenges include local resource manager scalability and granularity, efficient utilization of the raw hardware, shared file system contention and scalability, reliability at scale, application scalability, and understanding the limitations of the HPC systems in order to identify promising and scientifically valuable MTC applications. This paper presented new research, implementations, and application experiences in scaling loosely coupled applications on the Blue Gene/P up to 128K processors and microbenchmarks up to 160K processors.

### A. Characterizing MTC Applications for Petascale Systems

Based on our experience with the Blue Gene/P at 160K CPU scale (nearly 0.5 petaflop Rpeak) and its shared file system (GPFS, rated at 8GB/s), we identify the following characteristics that define MTC applications that are most suitable for peta-scale systems:

- *Number of tasks >> number of CPUs*
- *Average task execution time > O(60 sec) with minimal I/O to achieve 90%+ efficiency*
- *1 second of compute per processor core per 5KB~50KB of I/O to achieve 90%+ efficiency*

The main bottleneck we found was the shared file system. GPFS is used throughout our system, from booting the compute nodes and I/O nodes, to starting the Falkon dispatcher and executors, starting the applications, and reading and writing data for the applications. Assuming a large enough application, the startup costs (e.g. 1326 seconds to bootstrap and be ready to process the first task at 160K processors) can be amortized to an insignificant value. We offloaded the shared file system to in-memory operations by caching the Falkon middleware, the applications binaries, and the static input data needed by the applications in memory, so repeated use could be handled completely from memory. We found that the three applications we worked with all had poor write access patterns, in which many small line-buffered writes in the range of hundreds of bytes were performed throughout the task execution. When 160K CPUs are all doing these small I/O calls concurrently, it can slow down the shared file system to a crawl, or, even worse, crash it. The solution was to read dynamic input data from shared file system into memory in bulk (e.g., dd with block sizes of 128KB), let applications interact with their input and output files directly in memory, and write dynamic output data from memory to shared file system in bulk (e.g., dd, merge many output files into a single tar archive).

### B. Future Work

Many MTC applications read and write large amounts of data. To support such workloads, we want to make better use of the specialized networks found on some petascale systems, such as the Blue Gene/P's Torus network. Our efforts will in large part focus on having transparent data management solutions to offload the use of shared file system resources when local file systems can handle the scale of data involved.

One solution is to exploit unique I/O subsystem capabilities of petascale systems. For example, collective I/O operations could be implemented to use the specialized high bandwidth and low latency interconnects, such as the Torus network. Through supporting collective I/O operations, we hope to be able to support more efficient and scalable solution for common file access patterns, such as broadcasting a common file across to all compute nodes or aggregating many unique files from many compute nodes into few files that can be written to GPFS with relatively small number of I/O calls and with little to no contention on GPFS.

We expect that data caching, proactive data replication, and data-aware scheduling will offer significant performance improvements for applications that exhibit locality in their data access patterns [35]. We have already implemented a data-aware scheduler, and support for caching in the Falkon Java executor, under the umbrella of data diffusion [34, 35, 8]. In previous work, we have shown that in both microbenchmarks and a large-scale astronomy application, a modest Linux cluster (128 CPUs) can achieve aggregate I/O data rates of tens of gigabytes of I/O throughput [34, 8]. We plan to port the same data caching mechanisms from the Java executor to the C executor so we can use these techniques on the Blue Gene/P by leveraging the Torus network interconnect to communicate directly between compute nodes. We have already completed the first step towards this goal, to enable TCP/IP connectivity over MPI of the Torus network which gives us a global IP space among all compute nodes, as opposed to the private IP space per pset that we had using the Tree network.

MTC applications could also be composed of individual tasks that are themselves parallel programs. We plan to add support for MPI-based applications in Falkon, specifically the ability to run MPI applications on an arbitrary number of processors. We have a candidate application that needs to have thousands of separate MPI-based application invocations, with each invocation getting optimal performance with 32 processors. This use case is one that is not well supported today on the Blue Gene/P because MPI applications currently have to use processors in pset granularity (256 processors).

## Acknowledgment

We thank the Argonne Leadership Computing Facility for hosting the IBM Blue Gene/P experiments. We also thank our colleagues for valuable help and feedback working with the DOCK and MARS applications, namely Don Hanson, Rick Stevens, Matthew Cohoon, and Fangfang Xia. Special thanks are due to Mike Kubal for his work on, and explanation of, the molecular dynamics applications.